\documentstyle[prl,aps,epsfig]{revtex}
\begin{document}
\title{Observing Quanta on a Cosmic Scale}
\author{Craig J. Hogan}
\address{Astronomy and Physics Departments, 
University of Washington,
Seattle, Washington 98195-1580}
\maketitle
\begin{abstract}
Our entire galaxy, like all others, originated as a fairly smooth patch of binding energy, which in turn originated
as  a single quantum perturbation of the inflaton field on a subatomic scale during inflation.  The best preserved
relic of these perturbations is  the anisotropy of the microwave background radiation, which on the largest
scales preserves a faithful image of the primordial quantum fields. It is possible that close study of these
perturbations might reveal signs of discreteness caused by spacetime quantization. 
\end{abstract}

\section{cosmic background Anisotropy: Images of Inflaton Quanta on the Sky}
At the beginning of the first session of the Humboldt Foundation's  symposium 
 ``Werner Heisenberg: 100 years Works and Impact'', an eminent particle theorist 
  chose  to begin   the conference with a talk not on 
the grand tradition of Heisenberg quantum mechanics we had
gathered to celebrate,  but on the  much less elegant topic  of
 astrophysical cosmology.  This choice made perfect sense: at this moment in the development of science, the most
fundamental  advances in understanding the innermost world of small things are coming from studies of big things:
The newest  physics is coming from astrophysics.

The last few years have seen  spectacular  discoveries of this kind. Astrophysical experiments have determined that
neutrinos have mass,  demonstrating the first new extension of Standard Model phenomena in many years.
The mass-energy of the universe appears to be dominated by a new form of ``Dark Energy'' with a sufficiently large
repulsive gravitational  effect to accelerate the cosmic expansion of the universe, as detected in measurements of
distant supernovae, and which defies all attempts at a deep theoretical understanding.
 Another  great advance,
the detailed measurement of the pattern of primordial anistropy in the cosmic microwave background, has now
started to nail down basic parameters of the universe with some  precision.  We now know for example that the large
scale geometry of the universe is nearly flat, or to put it another way, the entire cosmic hypersphere appears to be
at least ten times larger in linear scale than the piece accessible to our telescopes. Better data soon to come will
probably drive that lower limit up to a factor of a hundred. Real data are
 confirming the expectation, from inflation
theory, that the universe is much larger than what we can ever see.

From the point of view of Heisenberg, perhaps    most amazing newly discovered phenomena are the perturbations that
create   the anisotropy. Hot and cold patches in the cosmic background radiation correspond to small fractional 
perturbations in gravitational potential,    on a vast scale, with coherent patches stretching across the entire
visible universe. The perturbations are also ultimately responsible for causing all of the structure in the universe,
the superclusters of galaxies, and the galaxies, stars and planets within them. They are most remarkable however
because they are a quantum phenomenon: the pattern of hot and cold patches scientists map  on the sky today is a
faithful image of the configuration of  quantum field fluctuations  that occurred in a tiny patch of space, far
smaller than an atom, long ago during cosmic inflation. Roughly speaking, each hot and cold patch originated as  a
single quantum of the inflaton field, which was subsequently enormously magnified by the cosmic expansion acting as a
coherent, nearly noise-free amplifier.

This    audacious application of quantum mechanics is generally accepted as the correct explanation
 of the
origin of these perturbations because a well-controlled  
model\cite{Starobinsky:1979ty,bardeen1980,Hawking:1982cz,Guth:1982ec,Bardeen:1983qw,Starobinsky:1982ee,Halliwell:1985eu,Grishchuk:1993ds}
 makes definite predictions that agree well with the
data\cite{cobeDMR,boom,pryke,halverson,max,bond01}. The basic framework of the theory consists of free relativistic
quantum fields in a classical curved spacetime: that is, Einstein's classical theory of gravity is used to compute the
spacetime background on which the quantum fields propagate. The fields affect the spacetime through  their
energy-momentum tensor, as described by the Einstein field equations, but the spacetime itself is not quantized. The
fields approximate a collection of quantized harmonic oscillators in an expanding background, with perturbations due
to the zero-point field amplitudes in their ground state. Each mode expands  with the background universe, and at a
certain point its fluctuations are ``frozen in'' as its oscillation rate becomes smaller  than the expansion 
rate, a process
that can most
accurately be
descibed as a phase
wrapping or state
squeezing\cite{Grishchuk:1989ss,Grishchuk:1990bj,Albrecht:1994kf,Lesgourgues:1997jc,Polarski:1996jg,kiefer}. The
fluctuation becomes a perturbation in the spacetime metric\cite{bardeen1980},  which continues to expand by a huge
factor as the universe exponentially increases in size during inflation. The final superposition of all the modes
creates metric perturbations described as a continuous field with  random Gaussian statistics, and a nearly (though
not exactly) scale-free spectrum. These predictions are confirmed by the current
data.

Thus the  structure of the universe on the largest scales is directly connected to quantum processes on the
smallest scales.
The statistics of the fluctuations, such as the amplitude and slope of their spectrum,
 are determined by certain combinations of  parameters of the field Lagrangian. The
data on anisotropy  provide  
 by far the most precise data we have   on the structure of fundamental forces on such small scales
(albeit, the information is limited in scope to a few special combinations of parameters).

Indeed, the theory extrapolates the basic theoretical framework 
 tens of orders of magnitude  from any other experimental data, which leads to some healthy scepticism 
about whether we ought to believe it at all.  In some situations (in particular, if inflation occurs close enough to
the Planck scale to produce detectable and separable tensor perturbations due to graviton fluctuations, in addition to
the scalar modes that lead to galaxies),  we may be able to implement  a more detailed test, for example, a comparison
of the relative amplitude of scalar and tensor mode perturbations with the slope of the
spectrum\cite{Mukhanov:1992me,lythriotto}.  However, for plenty of choices of parameters    the
field theory framework predicts only unobservable departures from
 scale-invariant, random Gaussian noise.

\section{  holographic information bound and quantum discreteness of anisotropy}

It has always been acknowledged that standard  inflation theory is  only itself an approximation, because it does not
include quantization of spacetime itself--- and for good reason, since there is not yet a widely accepted  theory of
quantum gravity. On the other hand, there are now some definite quantitative results in quantum gravity:   the
dimension of the Hilbert  space is known or bounded, giving   bounds on the total entropy and information accessible
to all configurations of all fields, including the quantum degrees of freedom of the spacetime itself. Instead of
having an arbitrary zero point, entropy can now be  defined in absolute terms. In other words,
there are absolute limits on how many different things can happen within the confines of any given region. 

 This in turn imposes bounds on
everything during inflation, including the behavior of free quantum fields. Their modes are not truly and
fundamentally independent as assumed in the standard picture, nor do they have an infinite Hilbert space. This is a
radical departure from the foundations of field theory, but it is required if we wish to include spacetime as a
quantum  object in the fundamental theory, in particular one that obeys unitary evolution without fundamental loss of
information.

Consider for example the thought experiment of a black hole that forms, then evaporates via Hawking 
radiation
\cite{Bekenstein:1972tm,Bekenstein:1973ur,Bekenstein:1974ax,Hawking:1975sw,Hawking:1976ra,Bekenstein:1981jp}.  If
the whole process is unitary\cite{Banks:1984by,'tHooft:1985re,Susskind:1993if,stephens}, then the states of radiated
particles depend in detail on how  the hole was formed. Indeed, running things backwards in time, carefully
assembled\footnote{To do this right with the known CP-violating, CPT-invariant fields,
one would have to take the final states and parity reverse them before running them backwards in time.} incoming
particles would form a small black hole that grows and then disassembles by throwing out any particular macroscopic
objects--- TV sets, whatever--- that  went into the hole. This  is only possible if the spacetime metric encodes at a
fundamental level the information equivalent to the radiated entropy.  The entropy is known from thermodynamic
arguments to be one quarter of the area of the event horizon in Planck units. Since a black hole is   the highest
entropy state attainable by any amount of mass/energy,   one is led to  the ``Holographic Principle''
\cite{thooft93,susskind95,Bigatti:1999dp}: for any physical system,
   the total entropy $S$
   within any surface is  bounded by one quarter  of the area $A$ of the surface in
Planck units  (adopting $\hbar=c=G=1$). This is an
``absolute'' entropy; the dimension
$N
$ of the Hilbert space  is given by $e^S$, and the total number of distinguishable quantum states 
available to the system is given by a binary number with $n=S/\ln 2=A/4\ln 2$
 digits \cite{bekenstein01}.

A holographic bound can also be derived for the universe   as a
whole\cite{susskindwitten,fischler99,bousso,Bousso:1999cb,Bousso:2000dw,bousso99}. Because the cosmological solutions
do not have the asymptotically flat infinity of the black hole  solutions for defining particle states, the bound has
a somewhat different operational meaning, referring to an ``observable entropy''.
 The  cosmological version\cite{Nbound,flanagan} of the holographic principle is:   the
observable entropy of any universe cannot exceed $S_{max}= 3\pi/\Lambda$,
where
$\Lambda$ is the cosmological constant in Planck units.   In a de Sitter
universe, as in a black hole, this corresponds to  one quarter of the area of
the event horizon in Planck units, but the bound is conjectured to hold for any
spacetime, even Friedmann-Robertson-Walker  universes with matter as well as 
$\Lambda$.    The inflationary part of our 
spacetime   closely resembles a piece of a  de Sitter universe, so there is a bound on the observable entropy of all
quantum fields during inflation:
$S_{max}=\pi/H_{}^2$, where $H$ is the   expansion rate during inflation.
That means that all the modes of all the fields have the same size Hilbert space as a system of 
$n=\pi/H^2\ln 2$ binary spins.

The value of $H$ during inflation is not known, but the field theory   predicts that  
  graviton fluctuations are produced with amplitude $\approx H$, leading to tensor perturbations of the metric and
anisotropy with amplitude $\delta T/T\approx H$; since the observations with
$\delta T/T\approx 10^{-5}$ are  dominated by scalar perturbations, we currently have a bound around $H\le 10^{-5}$.
(This bound will improve with the  advent of experiments  with better sensitivity to polarization that can separate
tensor and scalar components\cite{kamionkowski}). In round numbers then, the universe during inflation has a Hilbert
space equivalent to at least
$10^{10}$ spins. That is unquestionably a large number, but then again it is much smaller than the infinite Hilbert
space of the field theory description. It is also much smaller than the Hilbert space of
other astrophysical systems, such as stellar-mass black holes.

The most interesting  question is, can we detect any observable effect of the finite Hilbert space on the 
fluctuations? We derive some hope from the fact the the microwave background experiments have such a high precision,
and that the effects of various complicated astrophysical foregrounds have  been successfully removed to reveal the
simple, cleanly modeled perturbations in the cosmic last scattering surface.

To get one very rough estimate of  the possible amplitude of the effect, consider the following crude model. Pay
attention only to one set of modes, those five independent components
 that contribute to the observed quadropole moment of the
anisotropy. These perturbations are created by (and are amplified images of) quantum fluctuations around the time that
the current Hubble length passed through the inflationary event horizon; the amplitude of the observed temperature
perturbation traces in detail a quantity which is proportional to the amplitude of the quantum field. Imagine that
these modes contain all of the information allowed by the cosmological holographic bound, and further that they are
literally composed of $n$   pixels on the sky, each of which represents a binary spin. For the maximal value
$H\approx 10^{-5}$ (the maximum allowed by the tensor-mode limits), there would be of order $10^{10}$
``black-and-white'' pixels, corresponding to a pixel scale of only a few arc seconds.\footnote{This level of 
graininess in the everyday world would be just below the level of detectability with the naked eye.} This level of
discreteness would not be observable in practical terms, for two reasons: the last scattering surface is much thicker
than this (which  smears out the contribution of many pixels due to optical transfer and acoustic  effects  at a
redshift of about 1000), and there are many other modes on a smaller scale superimposed.

On the other hand, at least one important assumption in this estimate is likely to  be  wrong by orders of magnitude:
the modes on the horizon scale probably carry only a small fraction of the holographic information bound. At any given
time,  most of the information is in much smaller wavelength modes. A toy model of spacetime
discreteness\cite{hogan2001} suggests that the Hilbert space of the horizon-scale quantum perturbations is equivalent
to at most only about 
$10^5$ binary spins. (This number is determined not the tensor-mode limit on $H$,
but by the inverse of the observed scalar
perturbation amplitude). This in principle may produce observable discreteness, since the all-sky cosmic background
anisotropy includes about
$10^4$ independent samplings of  inflationary fields (determined by the angular size of the horizon at last
scattering).  

Now the idea that there are literally binary pixels is also silly, and was used here only  for illustrative
convenience. In fact we have no clear idea of  the actual character of the holographic eigenstates  projected onto
the inflaton perturbations; we have only the counting argument to guide us. In \cite{hogan2001}, 
a toy model assumed that
$H$ and $\delta T$ come in discrete levels, but this again was only for calculational convenience. In the absence of a
more concrete theory  of  the nature of the holographic modes, it makes sense to consider a variety of tests on the
data to seek departures from the field-theory prediction of a continuous, random Gaussian  field.
The most straightforward check will  be to 
 test whether the amplitudes of the harmonics $A_{\ell m}$ come in discrete values rather than being selected from
a random continuous Gaussian distribution. Although this test could already be performed 
fairly reliably using the all-sky COBE satellite data\cite{bennett,gorski},
it will become much more powerful using
 the all-sky data from the
MAP (and later PLANCK) satellites, which will have much finer angular resolution capable of cleanly resolving modes
far below  the horizon scale at recombination (a level already reached in ground- and balloon-based experiments 
with limited sky coverage). The high quality of this data will let us search for true quantum-gravity effects.

\acknowledgements

I am grateful for useful conversations with J. Bardeen, R. Bousso, and R. Brandenberger.
 This work was supported by NSF grant AST-0098557 at the University of Washington.

%\end{document}

{}

\end{document}